\RequirePackage{fix-cm}
\documentclass[twocolumn,epjc3]{svjour3}
\smartqed  
\RequirePackage{graphicx}
\RequirePackage[caption=false]{subfig}
\RequirePackage{xcolor}
\usepackage{bm}
\usepackage{xspace}
\usepackage{multirow}
\usepackage{xcolor}
\usepackage[british]{babel}
\usepackage{hyphenat}
\usepackage{amsmath,amssymb}
\usepackage{cite}

\usepackage{graphicx} 

\usepackage{subfig}
\usepackage{float} 
\graphicspath{{figures/}} 
\usepackage{booktabs}
\usepackage[hidelinks]{hyperref}

\newcommand{\beq}{\begin{equation}}
\newcommand{\eeq}{\end{equation}}

\journalname{Eur. Phys. J. C}
\begin{document}

\title{Enhancing Neutrino Reconstruction in Water-Cherenkov Air Shower Arrays Using Multi-Photosensors 
}

\author{J.~Alvarez-Mu\~niz\thanksref{addr1}
        \and
        R. Colalillo\thanksref{addr4,addr5}
        \and
        R. Concei\c{c}\~ao\thanksref{addr2,addr3}
        \and        B.~S.~Gonz\'alez\thanksref{e1,addr2,addr3}
        \and
        V. M. Grieco\thanksref{addr5,addr6}
        \and
        F. Guarino\thanksref{addr4,addr5}
        \and
        M. Pimenta\thanksref{addr2,addr3}
        \and
        B. Tom\'e\thanksref{addr2,addr3}
        \and
        M.~Waqas\thanksref{addr4,addr5}
}

\thankstext{e1}{e-mail: borjasg@lip.pt}

\institute{
Instituto Galego de F\'\i{}sica de Altas Enerx\'\i{}as (IGFAE), Universidade de Santiago de Compostela, Santiago de Compostela, Spain \label{addr1}
\and
Universit\`a di Napoli “Federico II”, Dipartimento di Fisica “Ettore Pancini”, Napoli, Italy \label{addr4}
\and
Istituto Nazionale di Fisica Nucleare (INFN), Sezione di Napoli, Napoli, Italy \label{addr5}
\and
LIP - Laborat\'orio de Instrumenta\c{c}\~ao e F\'isica Experimental de Part\'iculas, Lisbon, Portugal \label{addr2}
\and
Departamento de F\'isica, Instituto Superior T\'{e}cnico, Universidade de Lisboa, Lisbon, Portugal \label{addr3}
\and
 Scuola Superiore Meridionale, Napoli, Italy \label{addr6}
}
\date{Received: date / Accepted: date}

\maketitle

\begin{abstract}
In this article, the potential of water Cherenkov detectors equipped with multi-PMT modules for background-free upward-going neutrino detection and improved direction reconstruction is demonstrated. By analyzing signal time traces with transformer-based models, significant improvements in angular resolution are achieved compared to previous designs with larger PMTs. These detectors enable the reconstruction of neutrino directions with resolutions of approximately $10^\circ$ in azimuth and $7^\circ$ in zenith for high-signal events, corresponding to an overall opening angle of approximately $10^\circ$. This design reduces saturation effects and enhances directional sensitivity, particularly for high-energy neutrinos. The results highlight the potential of WCD arrays as complementary tools for neutrino astronomy, particularly in the context of multimessenger observations of transient astrophysical sources. The nearly continuous operation and wide field of view of these detectors further enhance their suitability for real-time monitoring and alert generation.


\end{abstract}

%

\section{Introduction}
\label{sec:intro}

Neutrinos, being electrically neutral and weakly interacting, offer a unique window into the most extreme environments in the Universe. Unlike charged cosmic rays, which are deflected by magnetic fields, neutrinos travel freely from their sources, making them ideal messengers for identifying the origins of high-energy astrophysical phenomena. The accurate reconstruction of the neutrino direction is therefore crucial for identifying transient sources such as gamma-ray bursts, active galactic nuclei, and neutron star mergers, which are expected to produce neutrinos in the GeV to PeV energy range \cite{Ahlers:2015lln,Ackermann:2022rqc,Kurahashi:2022utm,km3net2025PeV}.

Large water Cherenkov detector (WCD) arrays have shown high efficacy in high-energy gamma-ray detection \cite{HAWC,LHAASO,SWGO}. Their nearly continuous operation and wide field of view make them particularly well-suited for multimessenger astronomy \cite{Pimenta2018Astroparticle}. Moreover, in previous work~\cite{neutrinos_2024}, it was demonstrated that upward-going particle tracks could be identified in individual WCD stations, even when equipped with only one photosensor at the bottom and one at the top. This was achieved through the analysis of the PMT signal time traces, although with limited angular resolution. This work significantly advances those results by using multi-PMT modules, such as in various neutrino-dedicated experiments \cite{km3net_multipmt_2022,hyperk_multipmt_2019}, which provide enhanced directional sensitivity. 

Machine learning (ML) techniques have revolutionized data analysis in astroparticle physics, particularly for ground-based observatories, where they have significantly improved performance in tasks such as event reconstruction \cite{carrillo2019improving,GUILLEN201912,nieto2019ctlearn,erdmann2018deep} and gamma/hadron separation \cite{WCD_9SiPM_2020,alpha_2022,jonas_gnn_IACT_2023,gnn2024jonas,ghsep_2024_tibet,LHAASO_2020_gnn,HAWC2021ML}. Among these techniques, transformers, which are based on attention mechanisms \cite{attention_2017}, have emerged as state-of-the-art algorithms, often outperforming traditional convolutional neural networks (CNNs) \cite{footprint_2025,Watson2023HAWC}. In this work, transformer-based models are applied to the analysis of signal time traces recorded by photomultiplier tubes (PMTs), enabling a more precise determination of neutrino direction. This approach, combined with the use of multi-PMT modules, enables the identification of smaller solid angles in the sky, improving the detection of transient astrophysical sources and complementing dedicated neutrino observatories.

The paper is structured as follows. Section~\ref{sec:simulations} describes the simulation setup and detector configuration. Section~\ref{sec:analysis} presents the proposed method for angular reconstruction using a single WCD equipped with two multi-PMT modules. Section~\ref{sec:results} evaluates the performance of the method, and Section~\ref{sec:conclusions} presents the conclusions.

\section{Detector concept and simulations for angular reconstruction} \label{sec:simulations} 

The WCD unit in this work is based on the \textit{Mercedes} WCD design proposed in~\cite{neutrinos_2024,wcd2022mercedes}. The cylindrical detector retains its original dimensions (height $=1.7\,{\rm m}$, base area $\approx 12\,{\rm m^2}$) and features reflective interior walls to optimize Cherenkov light collection. Although the geometry remains unchanged, the photosensor configuration has been redesigned to improve the neutrino angular reconstruction performance.

In the previous design, two 8-inch PMTs were placed in the center of the upper and lower caps of the WCD (\textit{M1T1} detector). This PMT layout enabled direct detection of Cherenkov light pulses from particles entering the water volume, regardless of their direction. Although effective for neutrino detection, the azimuthal symmetry of this configuration limited its angular resolution, motivating the adoption of designs with more PMTs, such as the \textit{Mercedes} WCD with three PMTs at the bottom base and one at the top (\textit{M3T1} detector).

\begin{figure}[htb]
 \centering
\includegraphics[width=0.95\linewidth]{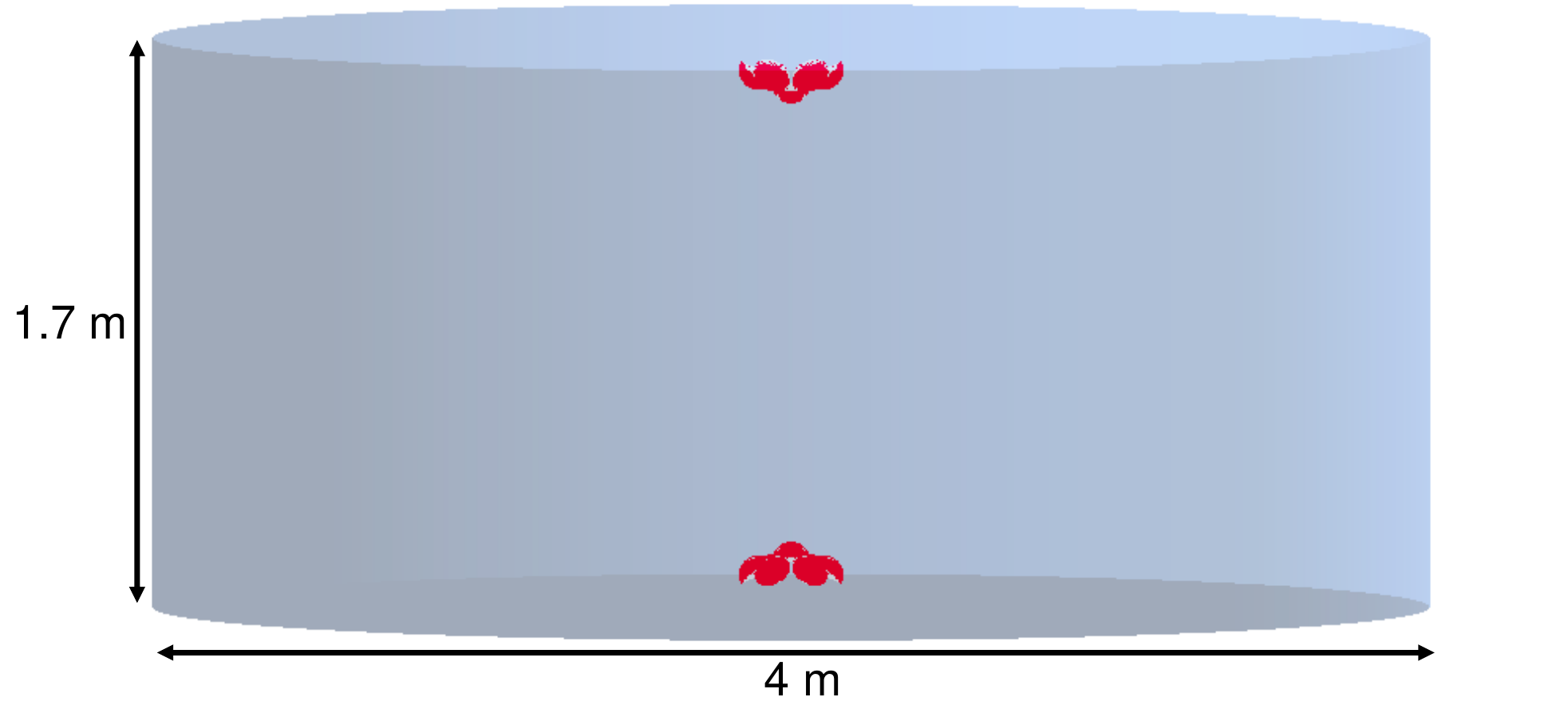}
\caption{\label{fig:M1mT1m} 
 \textit{M1mT1m} WCD design. The tank volume and the photocathode areas of the PMTs are represented in blue and red, respectively.}
\end{figure}

In this work, an upgraded design is explored, replacing the two PMTs with multi-PMT modules, each comprising seven 3-inch PMTs (\textit{M1mT1m detector}, see Figure \ref{fig:M1mT1m}). The PMTs are strategically arranged with a central vertical PMT and six peripherally distributed PMTs, tilted at $\theta_{\rm PMT} = 45^{\circ}$ in a hexagonal pattern. With a total photosensitive area that matches the original \textit{M1T1} design, this configuration ensures at least the same detection capability while providing azimuthal sensitivity, a critical improvement for angular reconstruction. Furthermore, multi-PMT modules offer redundancy compared to larger single-PMT configurations, helping to mitigate potential saturation effects in showers with large signals.

In addition, a cost-effective alternative to the \textit{M1mT1m} detector is explored, replacing the top multi-PMT module with a single 3-inch PMT (\textit{M1mT1s} detector). Since Geant4 simulations are very computationally demanding, this design was implemented by reusing the simulations of the \textit{M1mT1m} and excluding the six side PMTs of the upper multi-PMT module. The small PMT at the top might be particularly useful for recovering energy information in saturated stations, reducing systematic uncertainties in energy reconstruction, and improving gamma/hadron discrimination. Furthermore, its lower cost per WCD unit enables an increased array fill factor.

\begin{figure*}[htb]
 \centering
\includegraphics[width=0.95\linewidth]{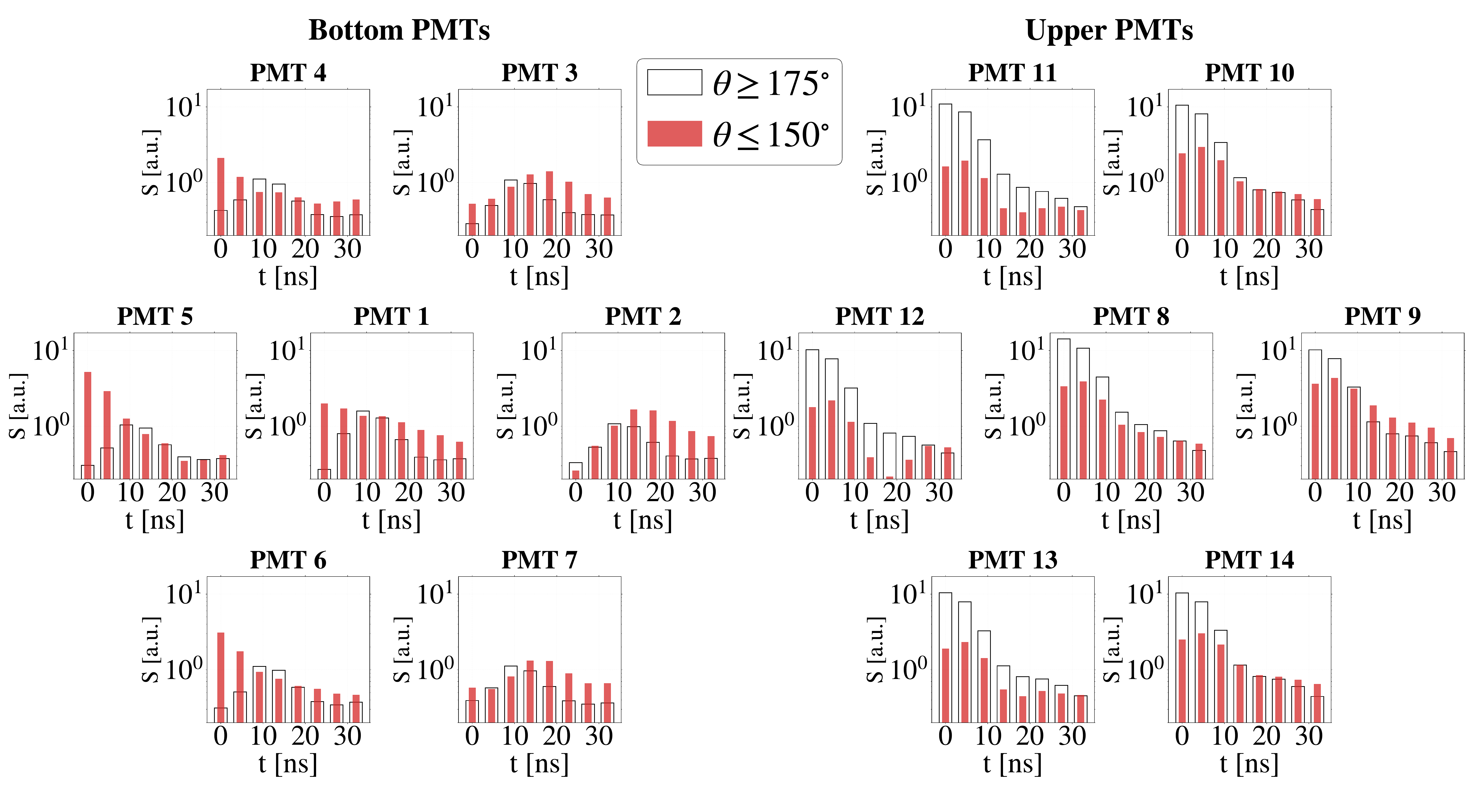}
\caption{\label{fig:mean_trace} 
 Preprocessed mean signal time trace for nearly vertical (black histogram: $\theta_0 \geq 175^{\circ}$, $\phi_0 \sim 180^{\circ}$) and inclined (red histogram: $\theta_0 \leq 150^{\circ}$, $\phi_0 \sim 180^{\circ}$) upward-going muons with an energy of $2 \, \rm GeV$. The events are centered within a radius of $r \leq 0.5 \, \rm m$. The waveform resulting from the PMT electronics simulation, originally in LSB units, has been preprocessed to determine $T_0$, the baseline of $402 \, \rm LSB$ has been subtracted, and it has been normalised by the gain $G=15 \, \rm LSB/p.e.$ The PMTs were arranged in a hexagonal layout to represent the real orientation of each PMT in the multi-PMT module.}
\end{figure*}

Upward-going muon neutrinos with an energy of $1 \, \rm TeV$ were simulated to evaluate the angular reconstruction capability of the detector. This energy corresponds to the intermediate energy range studied in \cite{neutrinos_2024}. The neutrinos were generated with a uniform distribution in azimuthal angle ($\phi \in [0, 360^\circ]$) and zenith angle ($\theta \in [135^\circ, 180^\circ]$), and at depths uniformly distributed within $z \in [-10, 0] \, \rm m$. To ensure that the neutrino interaction points are not restricted to the center of the detector, the $x$ and $y$ coordinates were shifted within a radius of 2 meters at the bottom base of the WCD. This adjustment allows the neutrinos to point to any location on the WCD bottom base. 

For this study, a total of roughly $150\,000$ upward-going $1 \, \rm TeV$ muon-neutrinos were simulated. Only events exceeding a signal threshold of $100$ photoelectrons (p.e.) were processed, a threshold determined based on the typical signal produced by vertically crossing muons in the WCD. This cut ensures that the analysis is focused on events in which particles fully cross the detector, while effectively suppressing low-energy background sources \cite{neutrinos_2024}. After applying this cut, the dataset was divided into $44\,107$ events for training, $11\,027$ for validation, and $69\,652$ for testing. The training set is used to optimize the model, the validation set for hyperparameter tuning and overfitting prevention, and the test set for unbiased performance evaluation.

A dedicated simulation was conducted following the approach outlined in~\cite{neutrinos_2024,wcd2022mercedes}. The detector response was simulated using Geant4 (version 4.10.05.p01)~\cite{agostinelli2003geant4,Geant4_2016,WCD4PMTs}, while the interactions of muon-neutrinos with the Earth's surface were simulated using HERWIG (version herwig6521)~\cite{HERWIG}. The secondary particles produced in these interactions were rotated to propagate upward, aligning their trajectories toward the WCD. The subsequent development of the neutrino-induced showers was simulated in Geant4, incorporating the ground density ($\rho=2.8\,{\rm{g\,cm^{-3}}}$) and material properties, accounting for all relevant physics processes in the shower.

The analysis in this work is based on the analysis of signal time traces, which are constructed through a detailed simulation of the PMT electronics chain. For each photoelectron, a pulse is generated using reference time and amplitude distributions, and the resulting analog waveform is formed by summing these individual pulses\footnote{The reference distributions for the PMT electronics simulation chain were taken from \cite{Electronics}.}. In addition, signal noise is included based on measurements from the experimental setup described in ~\cite{Electronics}. This noise was found to follow a Gaussian distribution with a standard deviation of approximately $\sigma \sim 0.8\,\mathrm{LSB}$, which is added to the analog signal before digitization.

The waveform is then digitized by sampling its amplitude at regular intervals, determined by a sampling rate of 250 mega samples per second (MSPS). To account for the arbitrary phase offset between the signal generation and the ADC sampling clock, the start of the digitization process is randomized within one sampling interval. This simulates the lack of synchronization with the master clock and introduces a realistic phase jitter in the digitized waveform. Specifically, a random time shift uniformly distributed in the interval $[0,\,1/f_\mathrm{sample}]$ is applied, where $f_\mathrm{sample}$ is the sampling frequency, which is $250\, \rm MSPS$. This approach reproduces the variations expected from unsynchronized sampling and enhances the realism of the simulation by introducing up to 4\,ns uncertainty in the alignment between the digitized signal and its true onset. The digitization process employs a 12-bit Analog-to-Digital Converter (ADC) with a range of $[0, 4095]$. In this step, the signal is truncated to the nearest integer and an offset of $400 \, \rm LSB$ is added. The simulation includes PMT saturation effects, with pulses exceeding $\sim\!250\,\mathrm{p.e.}$ being clipped by the 12-bit ADC's dynamic range.

To minimize the influence of electronics on the traces, a preprocessing step is applied after the PMT electronics simulation. First, a baseline of $402 \, \rm LSB$ is subtracted from the signals to account for the offset and Gaussian noise introduced by the electronics. Second, the $T_0$ of the event is determined by identifying the first time bin in which the cumulative signal across all PMTs exceeds a threshold of $5 \, \rm LSB$ above the baseline. This $T_0$ is then used to shift the time traces of all PMTs equally, aligning the event start to the beginning of the trace. In this way, the relative timing between signals in different PMTs is preserved, ensuring that temporal features, such as the time differences between peaks across PMTs, remain intact. Once $T_0$ is set, the signal is divided by the electronics gain, $G=15 \, \rm LSB/p.e.$ , and a time window of 32 ns after $T_0$ is used for further analysis, which is sufficient to capture the direct Cherenkov light and the first peak of reflected Cherenkov light \cite{wcd2022mercedes} (see Figure~\ref{fig:mean_trace}).

The output of the simulation of the PMT electronics chain, after the signal preprocessing step, is shown in Figure~\ref{fig:mean_trace}, which shows the mean signal time traces for nearly vertical ($\theta_0 \geq 175^{\circ}$) and inclined ($\theta_0 \leq 150^{\circ}$) upward-going muons with an energy of $2 \, \rm GeV$. The results demonstrate that this PMT layout effectively provides information about both the entry position of particles into the detector and their direction. For instance, vertical particles predominantly produce direct Cherenkov light in the top PMTs, while inclined particles can also generate direct Cherenkov light detectable in the bottom PMTs. This behavior is expected given the $41.2^\circ$ aperture of the Cherenkov light cone and the $45^\circ$ inclination of the peripheral PMTs, which optimizes their sensitivity to light from different angles.

\section{Neutrino effective mass}\label{sec:eff_mass}

To demonstrate the capability of the M1mT1m detector to discriminate particles induced by upward-going neutrinos from those produced by cosmic rays, the analysis pipeline developed in~\cite{neutrinos_2024}  was applied to this new detector configuration. 

The study relies on two sets of simulations. The first dataset is used to evaluate the background rejection capabilities of the detector. It includes background single particles—$1 \, \rm GeV$ electrons, $2 \, \rm GeV$ muons, and $10 \, \rm GeV$ protons—simulated at zenith angles ranging from $0^\circ$ to $92^\circ$, as well as events from proton-induced air showers with primary energies between $200 \, \rm GeV$ and $1 \, \rm TeV$, and upward-going electron and muon neutrinos with fixed energies of $10 \, \rm GeV$ and $1\, \rm TeV$. Since horizontal trajectories present the greatest challenge for discrimination, particular emphasis is placed on the angular interval $[87^\circ, 92^\circ]$, which accounts for half of the dataset's statistics. In total, this first dataset consists of $3 \cdot 10^5$ signal events and $2 \cdot 10^5$ background events, which are equally split into training and testing subsets.

The second dataset, used for the calculation of the neutrino effective mass, contains approximately $2.5 \cdot 10^5$ upward-going muon and electron neutrino events per simulation bin. These events are generated with fixed energies of $\{10 \, \rm GeV, 1 \, \rm TeV, 100 \, \rm TeV\}$ and various zenith angles: $\{0^\circ, 15^\circ, 30^\circ, 40^\circ\}$.

The identification chain begins with basic cuts on the signal time traces, designed to suppress background contamination. For simplicity, the summed response of the small PMTs in each multi-PMT module was used to approximate the performance of the single 8-inch PMT employed in the original M1T1 configuration. Although a dedicated set of cuts adapted to the geometry of the multi-PMT module could enhance performance, this lies beyond the scope of this work.

The applied selection criteria require: first, the maximum signal in the upper multi-PMT to occur in the first time bin, and second, the maximum in the lower multi-PMT to fall between the second and fourth bins. These conditions correspond to the direct detection of Cherenkov light followed by its reflection within the WCD, with a minimum time delay of $8 \, \rm ns$, based on the $4 \, \rm ns$ sampling rate and the $1.7 \, \rm m$ vertical height of the water volume. The chosen maximum positions are visible in the mean signal time traces of nearly vertical events in Figure~\ref{fig:mean_trace}. These signal cuts reject approximately $85\%$ of the background while preserving about $50\%$ of the signal events. These results are obtained under a conservative assumption for the detector electronics, where the PMT sampling rate is $250 \, \rm MSPS$ (corresponding to $4\, \rm ns$ time bins), matching the minimum time resolution required to apply the signal cuts effectively. Improved PMT electronics with higher sampling rates (i.e., thinner time bins) could potentially enhance the background rejection capability further by allowing a more precise characterization of the signal timing structure~\cite{neutrinos_2024}.

\begin{figure}[htb]
 \centering
\includegraphics[width=0.9\linewidth]{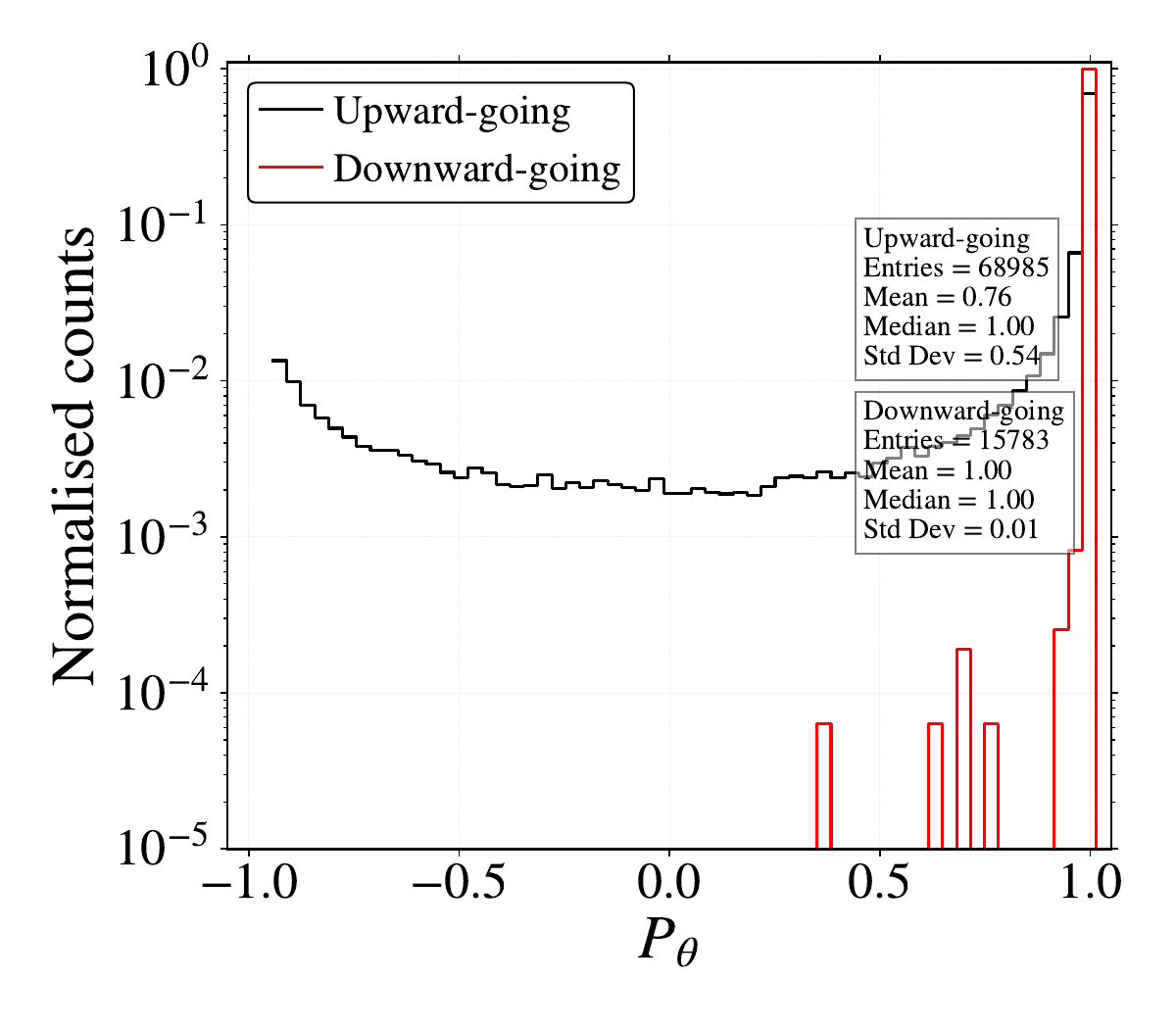}
\caption{\label{fig:CNN_M1mT1m} 
 Histogram of the CNN output for neutrino identification using the M1mT1m detector. Only the events passing the signal cuts were used to produce this plot.}
\end{figure}

In the final step of the analysis, a CNN is used to classify the surviving events. The CNN was trained to produce outputs of $-1$ for upward-going events (originating from negative $z$) and $+1$ for downward-going events (positive $z$). As shown in Figure~\ref{fig:CNN_M1mT1m}, all background particles yield CNN outputs above 0.5, confirming that none survive both the signal and CNN cuts. In contrast, a fraction of the neutrino-induced events are correctly identified, yielding outputs near $-1$.

To quantify the overall efficiency of the method, we estimate the neutrino effective mass for a WCD array composed of $5\,461$ M1mT1m detectors. The effective mass depends on the neutrino’s direction $(\theta, \phi)$, energy ($E_{\nu}$), the number of WCDs in the EAS array (\textit{N}), and interaction point $(x, y, D)$, where $D$ is the slant depth (in ${\rm g\,cm^{-2}}$) from the WCD station to the neutrino. Assuming a uniform distribution in $\phi$, the effective mass can be computed either for an isotropic flux using Equation\,(\ref{eq:neutrinos:dEff_mass}). 

\begin{equation} \label{eq:neutrinos:dEff_mass}
    \begin{aligned}
    M_{\text{eff}} \ (E_{\nu}) &= \int \dfrac{dM_{\text{eff}}}{d\theta} (\theta,E_{\nu}) \ {\rm d} \theta \\
    &= 2\pi  N \int \sin{\theta} \ \varepsilon(x,y,D,\theta,E_{\nu})\ {\rm d}x \ {\rm d}y \ {\rm d} D \ {\rm d}\theta \ {\rm [g]}
    \end{aligned}
\end{equation}

\begin{figure*}[htb]
 \centering
\includegraphics[width=0.9\linewidth]{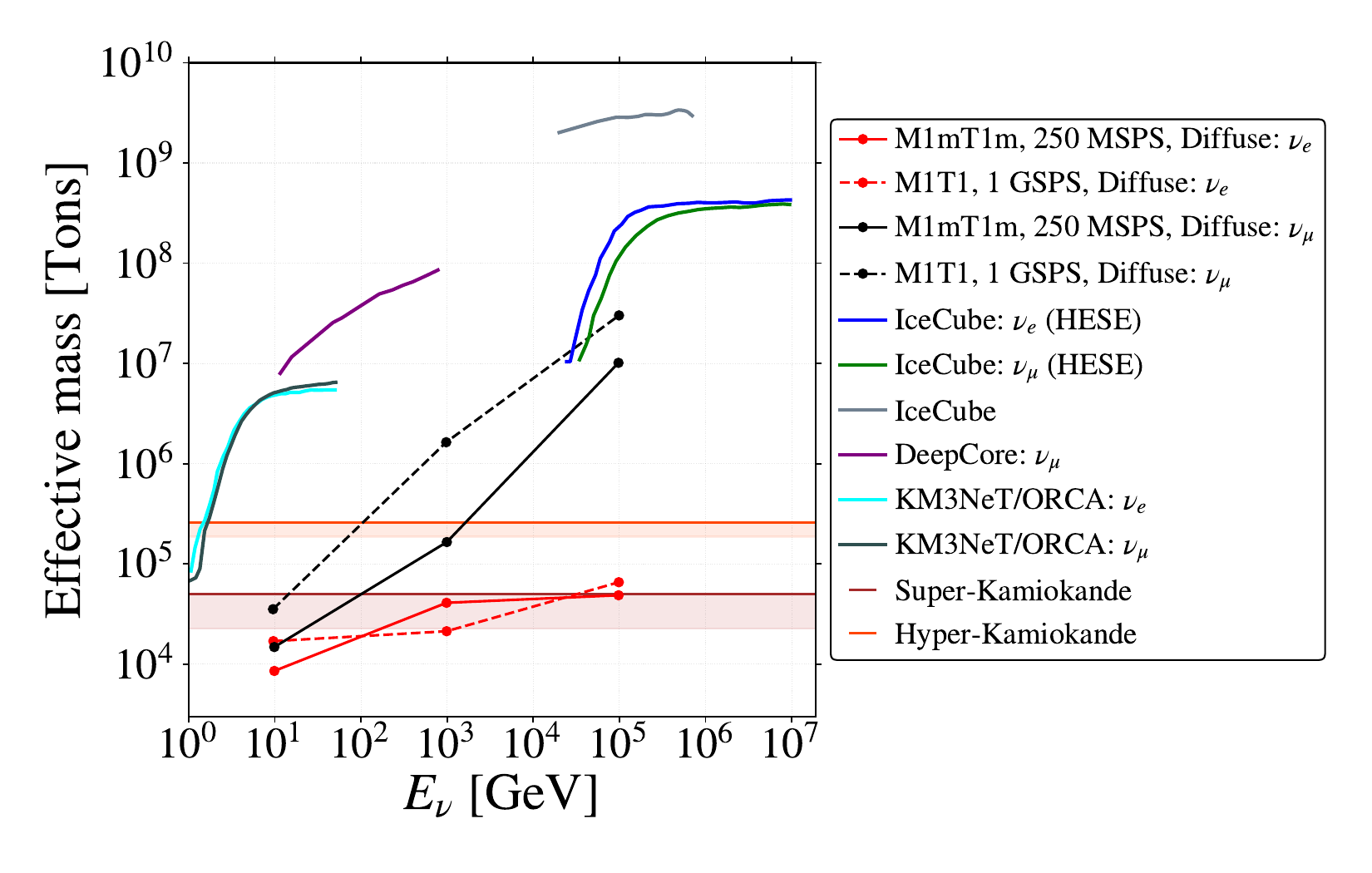}
\caption{\label{fig:Eff_mass_M1mT1m} 
 Neutrino effective mass obtained for a diffuse flux of neutrinos using the M1mT1m detector with a $250 \, \rm MSPS$  sampling rate ($4 \, \rm ns$ time bins), compared with the results from our previous study using the M1T1 detector with a $1 \, \rm GSPS$ sampling rate \cite{neutrinos_2024} ($1 \, \rm ns$ time bins) both using a EAS array with $5\,461$ WCDs. The results are compared with the effective masses of the following experiments: Super-Kamiokande \cite{SuperKamiokande}, Hyper-Kamiokande \cite{HyperKamiokande}, DeepCore online filter for $\nu_\mu$ (taken from Fig.\,2 of reference \cite{DeepCore2016}), KM3NeT/ORCA (taken from Fig.\,1 of reference \cite{KM3NeT2022EffMass}), IceCube (derived from Figures 2 and 5 of reference \cite{icecube2023observation}), and IceCube $\nu_e$ and $\nu_\mu$ High-Energy Starting Events (HESE) (taken from Fig.\,7B of reference \cite{icecube2013EffMass}). A shaded area between the fiducial and total water volume of the Kamiokande detectors was added.
}
\end{figure*}

In Figure~\ref{fig:Eff_mass_M1mT1m}, the resulting effective mass is presented and compared with the one of the M1T1 detector~\cite{neutrinos_2024}. This result demonstrates that the M1mT1m detector can provide effective background-free neutrino identification. For electron neutrinos, the effective mass is on the order of $10^4 \, \rm tons$, comparable to that of the Super-Kamiokande experiment. In the case of muon neutrinos, the effective mass increases more significantly with energy, as muons can traverse longer distances through the ground before reaching the detector. Specifically, it reaches approximately $10^4 \, \rm tons$ at $10 \, \rm GeV$—again comparable to Super-Kamiokande—scales up to $10^5 \, \rm tons$ at $1 \, \rm TeV$, similar to the water mass of Hyper-Kamiokande, and ultimately approaches $10^7 \, \rm tons$ at $100 \, \rm TeV$, closer to the performance of IceCube for High-Energy Starting Events.

The performance degradation compared to the M1T1 detector presented in~\cite{neutrinos_2024} is mainly attributed to the lower sampling rate used in the electronics simulation, rather than to the detector concept itself: $250 \, \rm MSPS$ for the M1mT1m configuration versus $1 \, \rm GSPS$ in the previous study. In particular, this reduced sampling rate has a significant impact on the efficiency of the signal cuts. While the $1 \, \rm GSPS$ setup achieved a background rejection rate of approximately $99\%$, similar signal cuts only reach about $85\%$ efficiency with the $250 \, \rm MSPS$ configuration adopted here.

This result underscores the critical role of the front-end electronics in enabling effective background rejection. In the present study, a conservative, worst-case scenario is considered, where the PMT sampling rate is equal to the time resolution required to apply the signal cuts described earlier. Improving the sampling rate could thus enhance the detector's performance in terms of background suppression.

\section{Methodology for angular reconstruction}\label{sec:analysis}
\begin{figure*}[htb]
 \centering
\includegraphics[width=1\linewidth]{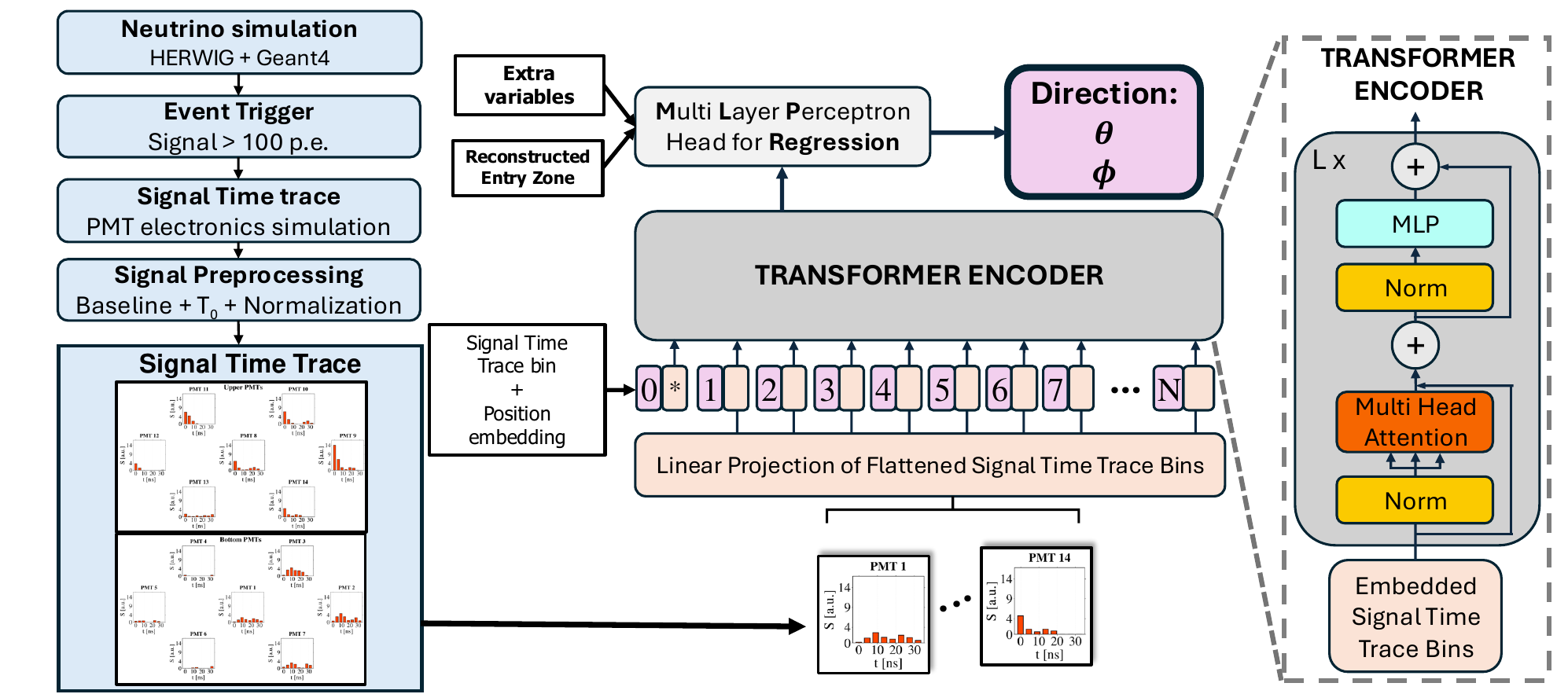}
\caption{Flowchart of the angular reconstruction pipeline using the Transformer. Adapted from \cite{ViT_2020}.}
\label{fig:flowchart} 
\end{figure*}

The reconstruction of the incoming neutrino direction is performed using the signal time traces from the PMTs. Although previous work used 1-dimensional CNNs for signal time trace analysis with similar WCDs \cite{neutrinos_2024,wcd2022mercedes,WCD4PMTs,4PMTs_NCA}, the increased complexity due to the higher number of PMTs in this setup required more advanced techniques. In this work, an innovative approach is employed, using a transformer encoder \cite{attention_2017} to extract features from the signal time traces, followed by a Multi-Layer Perceptron (MLP) for regression or classification. 

The transformer encoder processes the signal time traces by first splitting them into smaller segments, where each segment corresponds to the signal values of a single time-trace bin. Each segment is embedded into a higher-dimensional space, and positional encodings are added to retain temporal information. The segments are then processed through multiple transformer blocks, each consisting of self-attention and feedforward layers, to extract meaningful features. These features were subsequently used for the reconstruction of the neutrino direction.

In addition to the features extracted from the signal time traces using the transformer encoder, other engineered variables were incorporated into the MLP to enhance the reconstruction performance. These variables include: (1) total signal in the WCD; (2) the proportion of the total signal recorded by each PMT, normalized by the total signal across all PMTs; (3) the time bin corresponding to the maximum signal in each PMT; and (4) the absolute difference in time between the maximum signal in each PMT and the maximum signal in the opposite multi-PMT module. These variables, which were previously used for neutrino identification in \cite{neutrinos_2024}, provide additional contextual information on the spatial and temporal distribution of the signal.

The reconstruction process is then divided into two steps. First, given the large number of events that share the same direction but differ in vertex positions, an estimate of the entry point of the particles in the WCD is computed. This is achieved using a transformer model with a classification head at the end of the MLP. The entry points are categorized into six zones, each covering an angle of $60^\circ$ and centered around the position of the peripheral PMTs. For each event, the entry point $(x_0, y_0)$ at ground level is determined by extrapolating a straight line from the vertex point $(x_v, y_v, z_v)$ along the direction of the neutrino. The true entry zone is then calculated based on the $(x_0, y_0)$ coordinates of the entry point.

Once the entry zone is classified, the direction of the particle is reconstructed using a transformer model with a regression head. Both the zenith angle ($\theta$) and the azimuthal angle ($\phi$) are reconstructed simultaneously by the same network, as they are intrinsically related and contribute to the observed signal time trace patterns. 

The transformer models were implemented using PyTorch, with hyperparameters such as the number of layers, attention heads, and feedforward dimensions optimized using the Optuna framework \cite{optuna2019}. The optimal configuration included 2 transformer layers, 8 attention heads, a model dimension of 32, and a feedforward dimension of 128. Furthermore, the model featured two dense layers with 64 and 16 neurons, respectively, using ReLU activation and dropout rates of 0.1. The Adam optimizer \cite{kingma2014adam} was used with a learning rate of $0.001$, betas of $(0.9, 0.999)$, and a learning rate decay of $10^{-8}$. Training was conducted over $1\,000$ epochs with a batch size of $1\,024$ and an early stopping tolerance of 50 epochs to avoid overfitting. To ensure reproducibility and efficient monitoring of the training process, the results of the experiments were systematically tracked and managed using MLflow \cite{MLflow2018}.

For the classification of the entry zone, a standard cross-entropy loss was used. For the angular reconstruction task, on the other hand, the loss function was specifically designed to handle the periodic nature of angles, as proposed in~\cite{neutrinos_2024}. A circular regression loss function was used, defined in Equation \ref{eq:loss_function}, where $N$ is the number of samples, $y_i$ is the true angle for the $i$-th sample, and $\hat{y}_i$ is the predicted angle. This formulation ensures accurate reconstruction within the range of 0 to 360 degrees by explicitly accounting for the periodicity of angles. The loss is normalized by $1/2N$ to scale it between 0 and 1.

\begin{equation} \label{eq:loss_function}
{\rm Loss} = \frac{1}{2N} \sum^N_{i=1} \left[1 - \cos(y_i - \hat{y}_i)\right] \in [0,1] 
\end{equation}


In summary, the angular reconstruction process was illustrated in Figure~\ref{fig:flowchart}. The process begins with a trigger phase, during which only events exceeding the signal threshold of $100 \, \rm p.e.$ are processed, determined based on the typical muon signal. Once triggered, the signal time traces for each PMT are generated through the PMT electronics simulation chain. During preprocessing, a baseline of $400 \, \rm LSB$ is removed, $T_0$ is fixed to align the traces from different PMTs, and both the traces and engineered variables are standardized using z-score normalization. Subsequently, the particle's entry point is classified using a transformer-based model, followed by the reconstruction of the neutrino direction using another transformer model.

\section{Results} \label{sec:results}

\begin{figure*}[htb]
 \centering
    \subfloat[Histogram of the angular distance in $\phi$ between the true and predicted entry zones for neutrino events. An angular distance of zero denotes a correct classification of the entry zone.]{
   \label{fig:angular_error}
    \includegraphics[width=0.48\textwidth]{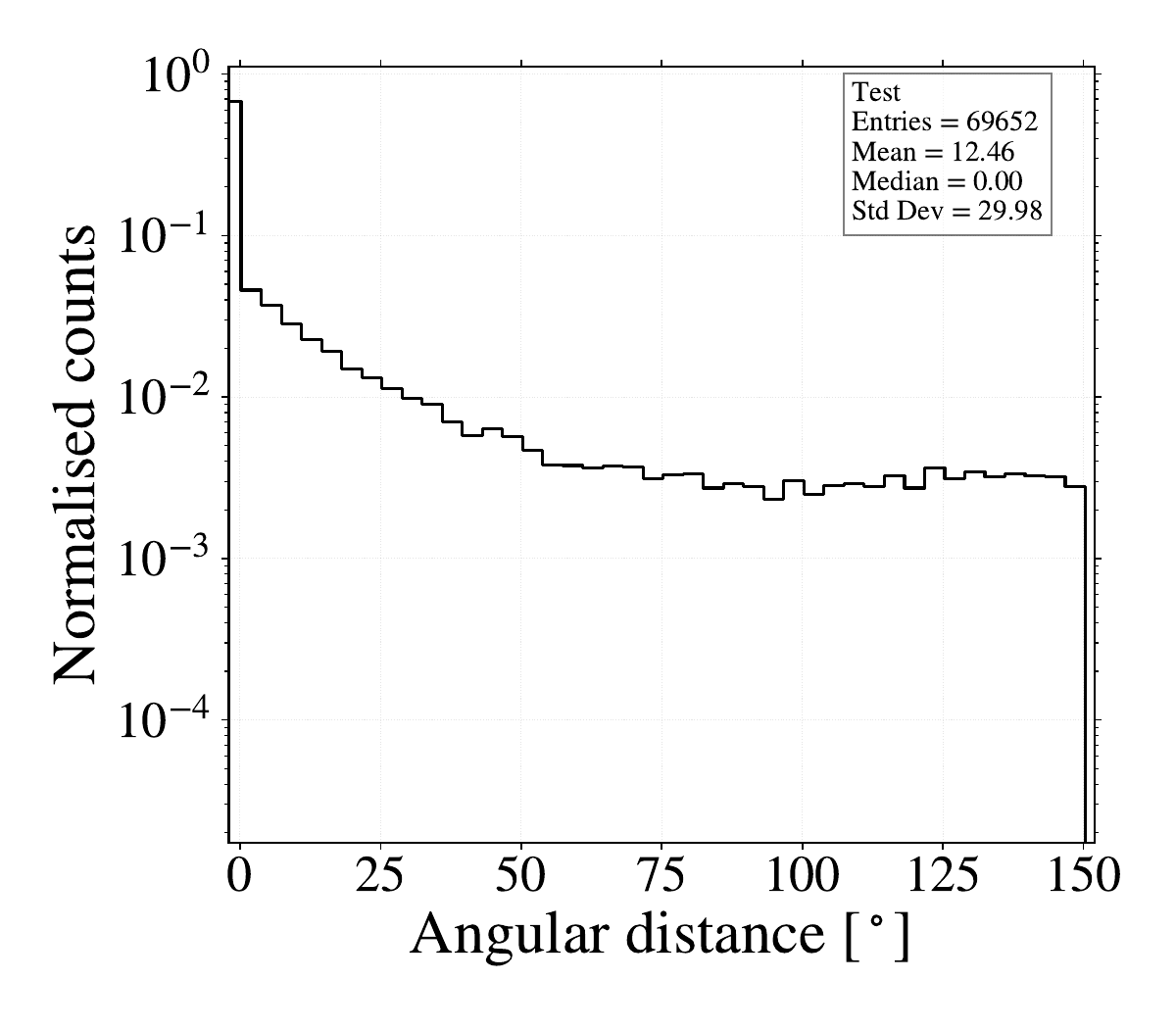}}
    \hspace{0.05in}
    \subfloat[Visualization of the entry points on the WCD surface, colored according to the reconstructed entry zone predicted by the transformer. The background colors represent the entry zones, which follow an anticlockwise order starting with \textit{Zone 0} centered at $0^\circ$ (red color).]{
 \label{fig:WCD_entry_point}
    \includegraphics[width=0.42\textwidth]{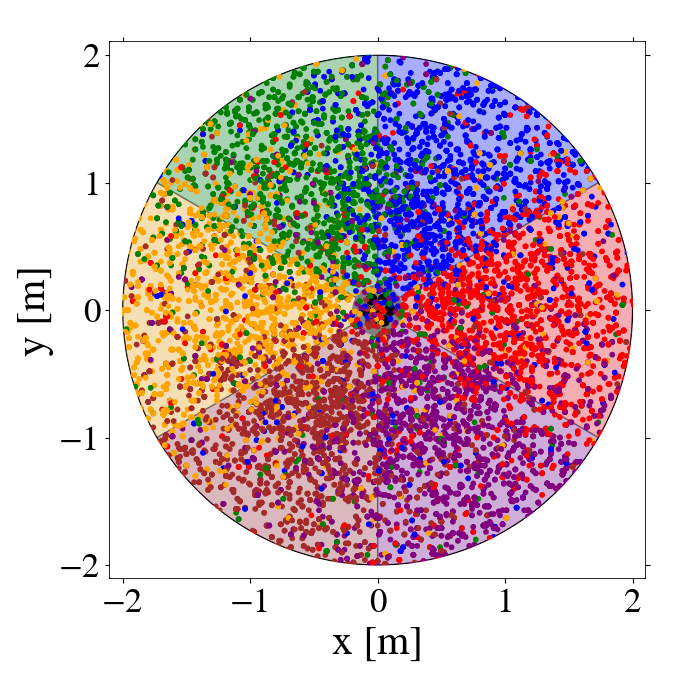}}
 \caption{
Performance of the entry zone reconstruction.
}
 \label{fig:performance_entry_point} 
\end{figure*}

As described above, the reconstruction process begins with the estimation of the entry zone. The performance of this step is depicted in Figure~\ref{fig:performance_entry_point}, which shows the distribution of the angular distance in $\phi$ between the true and predicted entry zones. The plot indicates that $\sim 70\%$ of the events are reconstructed with the correct entry zone position (angular distance of $0^\circ$), while most of the remaining misclassified events occur near the boundaries between zones, with angular distances lower than $30^\circ$. This is expected, as the Cherenkov cone angle of $41.2^\circ$ makes it difficult to classify events near the boundaries of the $60^\circ$ zones, where direct light can be detected by multiple PMTs.


It should be noted that, while the estimation of the entry zone is not essential for the final angular reconstruction, it serves as a valuable intermediate step. By providing additional context about the vertex position, the entry zone classification helps the transformer model discern patterns related to the neutrino direction, improving the overall reconstruction accuracy.

\begin{figure*}[htb]
 \centering
    \subfloat[Azimuthal angle.]{
   \label{fig:Phi_Reconstruction}
    \includegraphics[width=0.48\textwidth]{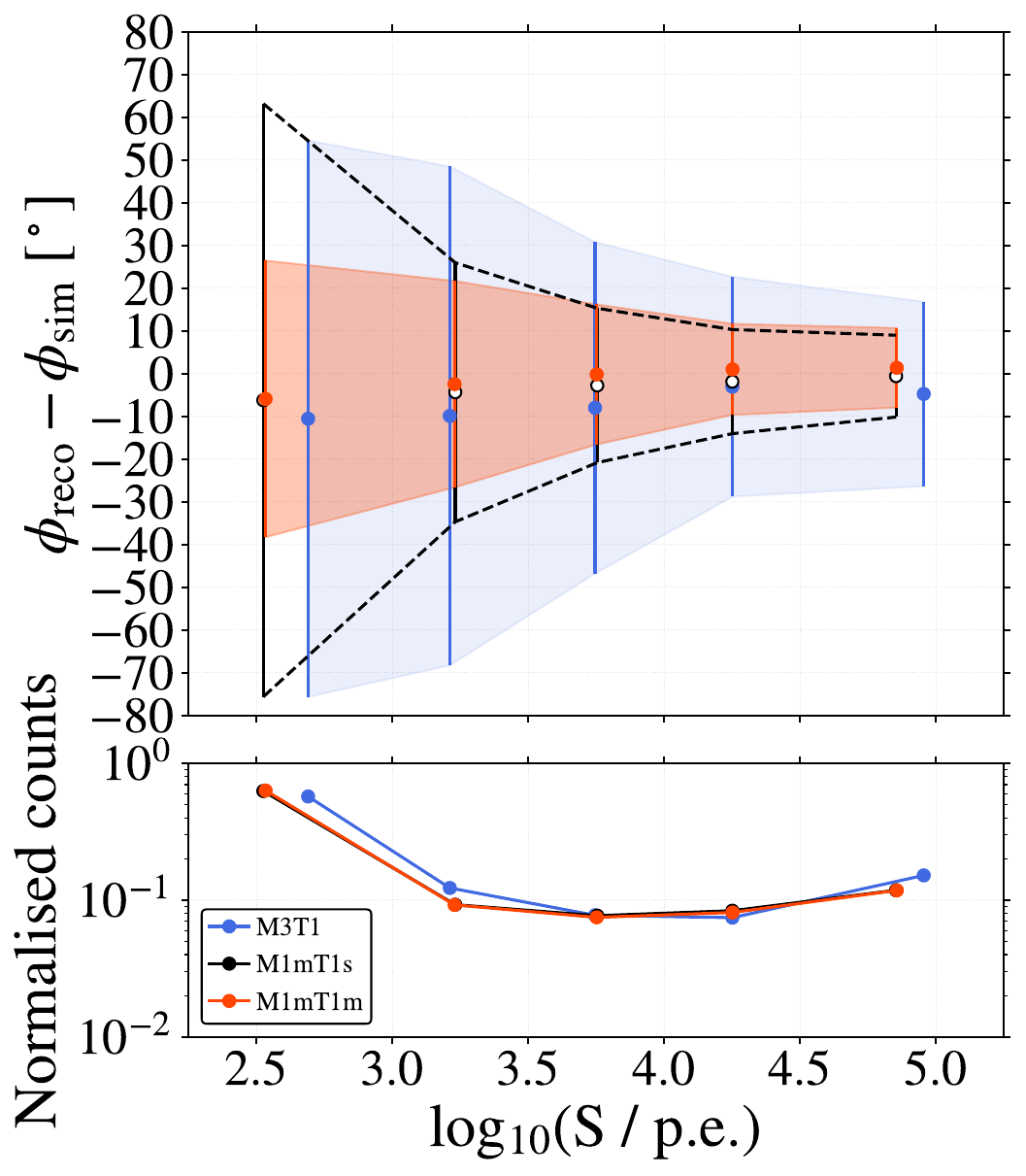}}
    \hspace{0.05in}
    \subfloat[Zenith angle.]{
 \label{fig:Theta_Reconstruction}
    \includegraphics[width=0.48\textwidth]{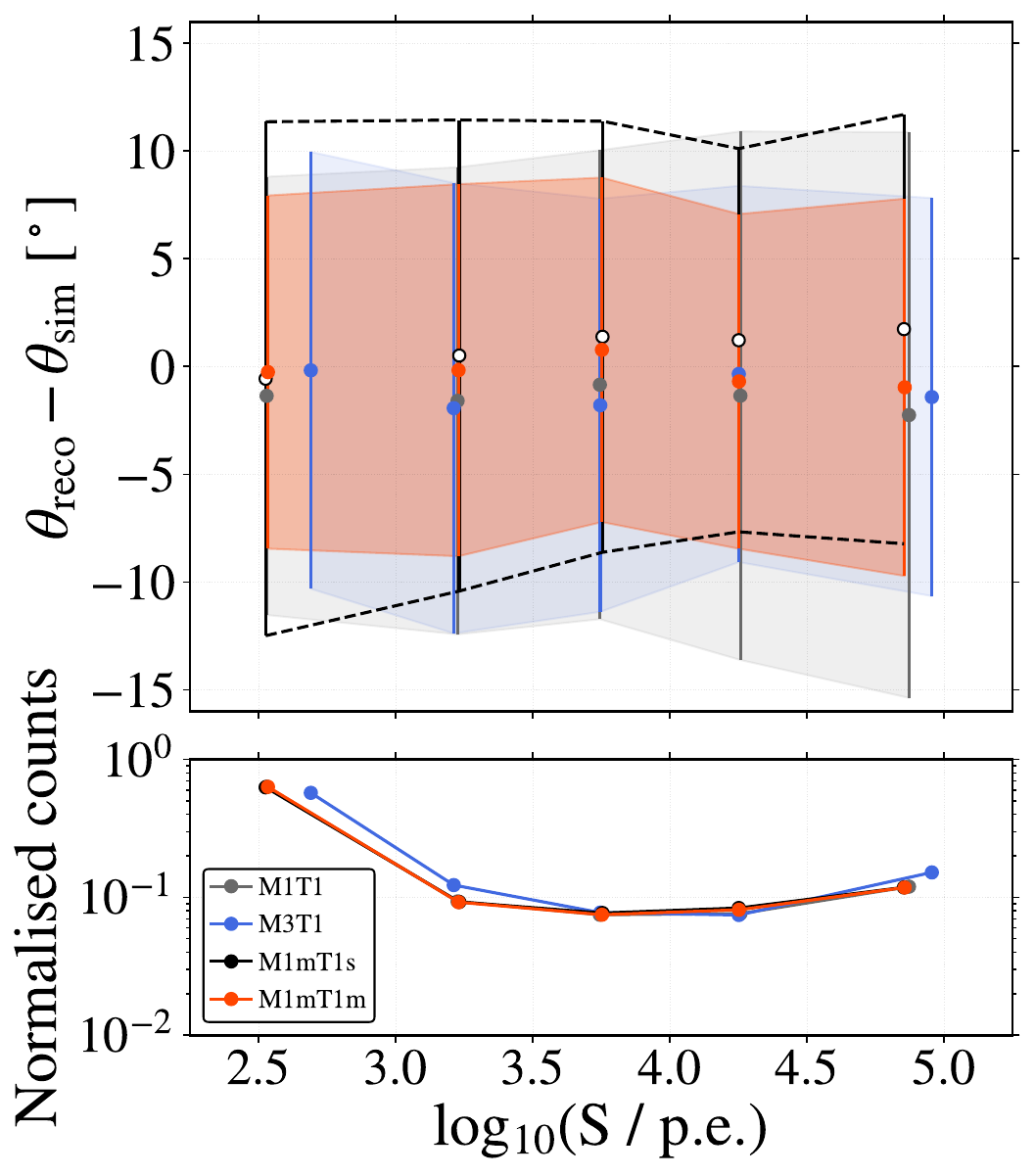}}
 \caption{
Comparison of the angular reconstruction efficiency with respect to the true neutrino direction as a function of the signal for the \textit{M1T1} (gray, \cite{neutrinos_2024}), \textit{M3T1} (blue, \cite{neutrinos_2024}), \textit{M1mT1s} (dashed black line, this work), and \textit{M1mT1m} (orange, this work) detectors. The top panel shows the angular reconstruction efficiency, while the bottom panel displays the number of events within each signal bin, normalized to the total number of events in the test dataset. Error bars represent the $68\%$ containment for each signal bin. Note that the \textit{M1T1} detector is not able to reconstruct $\phi$ due to the azimuthal symmetry of its PMTs.
}
 \label{fig:M3T1_vs_MultiPMT} 
\end{figure*}

The performance of the angular reconstruction is assessed in Figure~\ref{fig:M3T1_vs_MultiPMT}, which compared the efficiency of the WCD configurations using larger single PMTs, as described in previous work \cite{neutrinos_2024}, with the multi-PMT-based detectors introduced in this study. To ensure that performance differences arise solely from the detector concept and not from factors such as the reconstruction strategy or electronics, the same PMT electronics simulation and angular reconstruction strategy were applied consistently across all WCD designs. Furthermore, the hyperparameters of the transformer model were independently optimized for each WCD configuration using Optuna \cite{optuna2019}. The efficiency is quantified as the error between the true and reconstructed angles, plotted as a function of the total signal detected in the WCD. The bias of the distribution and the $68\%$ containment of the data are shown, representing the resolution of the method.

For events with a few hundred p.e., the reconstruction errors are approximately $30^\circ$ and $8^\circ$ for the azimuthal and zenith angles, respectively. As the signal increases to thousands of p.e., the resolution improves significantly, reaching approximately $10^\circ$ in $\phi$ and $7^\circ$ in $\theta$. The difference in signal is attributed to the vertex position: for low-signal events, the vertex is typically far from the WCD, resulting in only a muon crossing the detector. In contrast, high-signal events are likely to interact closer to the detector, allowing multiple particles from the neutrino-induced shower to cross the WCD and provide additional information for reconstruction.

The multi-PMT-based WCDs demonstrate a significant improvement over those using single larger PMTs, in particular for the reconstruction of the azimuthal angle. While the \textit{M1T1} is not sensitive to the azimuthal direction of the neutrino, the \textit{M1mT1m} design improves the \textit{M3T1} detector angular resolution by a factor of two, being $10^\circ$ for high-signal events. This enhancement is largely due to the distribution of the PMTs: in the \textit{M3T1} detector, the PMTs were positioned close to the WCD lateral wall in a $120^\circ$ star configuration, while in the \textit{M1mT1m} detector, the peripheral PMTs are separated by $60^\circ$. Moreover, the use of multiple smaller PMTs may reduce saturation effects, improving the reconstruction efficiency at the highest energies. It is important to highlight that, for events with $S \geq 1\,000 \, \rm p.e.$, the \textit{M1mT1s} exhibits a similar performance as the \textit{M1mT1m}. This result indicates that the upper multi-PMT module is essential to reconstruct low-signal events, effectively using direct Cherenkov light, while the patterns found on the reflected light captured on the bottom multi-PMT are enough to achieve an accurate angular reconstruction in high-signal events.

The resolution of the zenith angle with the \textit{M1mT1m} detector improves by only a few degrees compared to detectors based on single larger PMTs. This result is expected, as the total signal in the WCD--proportional to the muon track length--along with the time difference between the maximum signals in the upper and bottom PMTs and the time trace patterns, remain robust estimators across all detector designs. However, a smaller improvement was observed for the \textit{M1mT1s} configuration. This is primarily due to the reduced light collection efficiency of the single 3-inch PMT, which may miss direct Cherenkov light on the top PMT for certain particle trajectories. Furthermore, the side PMTs in the multi-PMT module are tilted by \(45^{\circ}\), providing additional information about the zenith angle.

To put these results in context, the opening angle between the reconstructed and true neutrino directions in this analysis is approximately $10^\circ$. In comparison, dedicated neutrino experiments such as IceCube and KM3NeT/ARCA achieve angular resolutions better than $1^\circ$ for track-like muon neutrino events in the TeV energy range. For cascade-like events, the resolutions are typically greater than $10^\circ$ for IceCube and around $2^\circ$–$3^\circ$ for KM3NeT/ARCA~\cite{km3net2025PeV,KM3NeT2019sensitivity,icecube2013EffMass,icecube2023observation}. Although the resolution of the \textit{M1mT1m} detector is less precise than dedicated experiments, it is important to note that this WCD was primarily designed for gamma-ray detection. Despite this, the achieved resolution is competitive, making the \textit{M1mT1m} detector a valuable complementary tool for neutrino astronomy, particularly for high-energy events where the neutrino flux is low and every detection is critical \cite{Ackermann:2022rqc,Kurahashi:2022utm}. It should also be noted that transient sources, such as gamma-ray bursts, active galactic nuclei, and neutron star mergers, are expected to produce neutrinos in bursts \cite{waxman1997high,ioka2005tev,abbasi2011neutrino}. In such cases, the temporal clustering of neutrino events, combined with their spatial coincidence, can provide strong evidence of their astrophysical origin, even if individual angular resolutions are limited \cite{icecube2013EffMass}. 

\section{Conclusions}\label{sec:conclusions}
In this work, we demonstrated the potential of WCDs equipped with multi-PMT modules for background-free neutrino detection and direction reconstruction, provided that the sampling rate of the PMT electronics is $250 \, \rm MSPS$ or better. The reconstruction is based on the analysis of signal time traces using transformer-based models. The multi-PMT-based WCD, featuring seven 3-inch PMTs arranged in a hexagonal pattern, provides enhanced directional sensitivity. This design enables the reconstruction of neutrino directions with angular resolutions of approximately $10^\circ$ in azimuth and $7^\circ$ in zenith for high-signal events, corresponding to a mean opening angle of about $10^\circ$ with respect to the true neutrino direction. This result represents a substantial improvement over results obtained with WCDs using multiple larger PMTs.

The angular reconstruction strategy with this WCD benefits from the finer azimuthal angular sampling provided by the $60^\circ$ separation of the peripheral PMTs. Compared to alternative WCDs that use multiple larger PMTs, the multi-PMT-based WCD improves azimuthal angle resolution by a factor of two and by a few degrees for the zenith angle, reducing the bias in both cases. In addition to the enhanced angular reconstruction, multi-PMTs mitigate the saturation effects by counting with more and smaller PMTs. This helps to recover energy information in saturated stations and may improve gamma/hadron discrimination.

A cost-effective alternative configuration was also evaluated, featuring a multi-PMT module at the bottom and a single 3-inch PMT at the top, demonstrating performance comparable to the dual multi-PMT design for high-signal events ($S \geq 1\,000 \, \rm p.e.$). 

The achieved angular resolution, while less precise than that of dedicated neutrino experiments such as IceCube and KM3NeT/ARCA, is competitive and suitable for identifying transient neutrino sources. This makes the multi-PMT-based WCDs a valuable complementary tool for neutrino astronomy, particularly in the context of multimessenger astronomy. The ability to reconstruct neutrino directions with high precision is especially important for high-energy events, where the neutrino flux is low and every detection is critical. Furthermore, the nearly continuous operation and wide field of view of WCD arrays make them ideal for real-time monitoring and alert generation, enhancing their role in the detection of transient astrophysical sources.

\section*{Acknowledgements}
We would like to thank Alberto Guillén for his valuable suggestions during the early stages of this work.
This work has been financed by national funds through FCT - Fundação para a Ciência e a Tecnologia, I.P., under project PTDC/FIS-PAR/4320.
B.S.G. (LIP/IST) is grateful for the financial support from the FCT PhD grant PRT/BD/151553/2021 under the IDPASC program (\url{https://doi.org/10.54499/PRT/BD/151553/2021}). J. A-M has received financial support from Ministerio de Ciencia, Innovaci\'on y Universidades/Agencia Estatal de Investigaci\'on, MICIU/AEI /10.13039/501100011033 Spain (PID2022-140510NB-I00, PCI2023-145952-2) and Mar\'\i a de Maeztu grant CEX2023-001318-M; Xunta de Galicia, Spain (CIGUS Network of Research Centers and Consolidaci\'on 2021 GRC GI-2033 ED431C-2021/22 and 2022 ED431F-2022/15).

\bibliography{references.bib}

\end{document}